\documentstyle{article}
\title{\bf Ideas Behind Kolmogorov Complexity\\
and Related Kolmogorov's Ideas about\\
The Physical Impossibility of\\
Large Integers:\\
How Their Formalization\\
Can Help in Foundations of Physics}

\author{Isaak A. Kunin$^1$, Vladik Kreinovich$^2$, and Yuri
A. Kuperin$^3$\\
$\ $\\
$^1$Department of Mechanical Engineering, University of Houston,\\
Houston, TX 77204, USA, email kunin@uh.edu\\
$^2$Department of Computer Science, Universuty of Texas at El Paso,\\
El Paso, TX 79968, USA, email vladik@cs.utep.edu\\
$^3$Physics Research Institute, St. Peterburg State University,\\
St. Petersburg, Russia, email kuperin@JK1454.spb.edu}
\date{}

\begin{document}
\maketitle
\begin{abstract}
In addition to the equations, physicists use the following 
additional difficult-to-formalize property: 
that the initial conditions and the value of the
parameters must not be abnormal. We will describe a natural
formalization of this property, and show that this formalization
in good accordance with theoretical physics.
At present, this formalization has been mainly applied to the
foundations of physics. However, potentially, more practical
applications are possible.
\end{abstract}

\section{Main Idea: In Short}

Traditional mathematical approach to the analysis of physical systems
implicitly assumed that all mathematically possible integers are
physically possible as well, and all mathematically possible
trajectories are physically possible. Traditionally, this approach has
worked well in physics, but it does not lead to a very good
understanding of chaotic systems, which, as is now known, are
extremely important in the study of real-world phenomena ranging from
weather to biological systems. 

  Kolmogorov was among the first who started, in the 1960s, analyzing
the discrepancy between the physical and the mathematical
possibility. He pinpointed two main reasons why a mathematical correct
solution to the corresponding system of differential or difference
equation can be not physically possible:
\begin{itemize}
\item First, there is a difference in understanding the term ``random" in
  mathematics and in physics. For example, in statistical physics, it
  is possible (probability is positive) that a kettle, when placed on a
  cold stove, will start boiling by itself. From the viewpoint of a
  working physicist, however, this is absolutely
  impossible. Similarly, a trajectory which requires a highly unprobable
  combination of initial conditions may be mathematically correct, but,
  from the physical viewpoint, it is impossible. 
\item Second, the traditional mathematical analysis tacitly assumes that
  all integers and all real numbers, no matter how large or how small,
  are physically possible. From the physical viewpoint, however, a
  number like $10^{10^{10}}$ is not physically possible at all, because it
  exceeds the number of particles in the Universe. In particular,
  solutions to the corresponding systems of differential equations which
  lead to some numbers may be mathematically correct, but they are
  physically meaningless. 
\end{itemize}
Attempts to formalizing these restrictions have been started by
Kolmogorov himself. These attempts are at present, mainly undertaken
by researchers in theoretical computer science who face a similar
problem of distinguishing between theoretically possible ``algorithms"
and feasible practical algorithms which can provide the results of
their computations in reasonable time. 

  The goal of the present research is to use the experience of
formalizing these notions in theoretical computer science to enhance
the formalization of similar constraints in working physics. 

  This research is mainly concentrated around the notion of Kolmogorov
complexity. This notion was introduced independently by several
people: Kolmogorov in Russia and Solomonoff and Chaitin in the
US. Kolmogorov used it to formalize the notion of a random
sequence. Probability theory describes most of the physicist intuition
in precise mathematical terms, but it does not allow us to tell
whether a given finite sequence of 0's and 1's is random or
not. Kolmogorov defined a complexity $K(x)$ of a binary sequence $x$ as
the shortest length of a program which produces this sequence. Thus, a
sequence consisting of all 0's or a sequence 010101\ldots  have a very
short Kolmogorov complexity because these sequences can be generated
by simple programs, while for a sequence of results of tossing a coin,
probably the shortest program is to write {\sf print(0101\ldots )} and then
reproduce the entire sequence. Thus, when $K(x)$ is approximately equal
to the length ${\rm len}(x)$ of a sequence, this sequence is random, otherwise
it is not. The best source for Kolmogorov complexity is a book
\cite{Li 1997}.

The definition of $K(x)$ only takes into consideration the length
${\rm len}(p)$ of a
program $p$. From the physical viewpoint, it is also important to take
into consideration its running time $t(p)$, because if it exceeds the
lifetime of the Universe, this algorithm makes no practical
sense. This development is in line with Kolmogorov's original
idea that some natural numbers which are mathematically possible (like
$10^{10^{10}}$) are not feasible and thus, should not considered as
feasible. Corresponing modifications are also described in the above
book. We plan to use the corresponding ideas in physics. 

\section{Main Idea: Brief Philosophical Analysis}
One of the main objectives of science is to provide {\it guaranteed}
estimates for physical quantities. In order to find out how estimates
can be guaranteed, let us recall how quantities are estimated in
physics:
\begin{itemize}
\item First, we must find a {\it physical law} that describes the
phenomena that we are analyzing. For some phenomena, we already know
the corresponding laws: we know Maxwell's equation for
electrodynamics, Einstein's equation for gravity, 
Schroedinger's equations for quantum mechanics, etc. (these laws can
be usually deduced from symmetry conditions 
\cite{Kreinovich 1976,Finkelstein 1985,Finkelstein 1986}). 
However, in many other cases, we must determine the equations from the
general theoretical ideas and from the
experimental data. Can we guarantee that these equations are correct?
If yes, how?
\begin{itemize}
\item [] There is an extra problem here. In some case, we know the
equations, but we are not sure about the values of the {\it
parameters} of these equations. If the theory predicts, e.g., that a
dimensionless parameter is 1, and the experiments confirm it with an
accuracy of 0.001, should we then use exactly 1 or $1\pm 0.001$ for a
guaranteed estimate? If the accuracy is good enough, then the 
physicists usually use 1. We may want to use $1\pm 0.001$ to be on the
safe side, but then, for other parameters of a more general theory (that
in this particular theory are equal to 0) should we also use their
experimental bounds instead of the exact 0 value? There are often many
possible generalizations, and if we take all of them into
consideration, we may end up with a very wide interval. This is a
particular case of the same problem: when (and how) 
can we {\it guarantee} that these are the right equations, with the
correct values of the parameters?
\end{itemize}
\item Suppose now that we know the correct equations. Then, we need to
describe how we will actually predict the value of the desired
quantity. For example,
we can get partial differential equations that describe how exactly
the initial values $\phi(x,t_0)$ of all the fields change in time.
Then, to predict the values of the physical quantity at a later moment
of time $t$, we must do the following:
\begin{itemize}
\item[$\bullet$]Determine the values $\phi(x,t_0)$ from the
measurement results. 
\item[$\bullet$]Use these values $\phi(x,t_0)$ to predict the desired
value.
\end{itemize}

\item[]The problem with this idea is that reconstructing the actual
values $\phi(x,t_0)$ from the results of measurements and observations
is an {\it ill-posed problem} \cite{TA77,I83,G84,I86,I86a,LRS86,CB86}
in the sense that two
essentially different functions $\phi(x,t_0)$ are consistent with the same
observations. For example, since all the measurement devices
are inertial and thus suppress the high frequencies, the functions
$\phi(x,t_0)$ and $\phi(x,t_0)+A\cdot\sin(\omega x)$, 
where $\omega$ is sufficiently
big, lead to almost similar values of observations. 
\begin{itemize}

\item[]A typical example
of an ill-posed problem is a problem of reconstructing the actual
brightness distribution of a celestial object 
from its observed image \cite{CB86}.
\end{itemize}

\item []Thus, strictly
speaking, if we do not have any additional 
restrictions on $\phi(x,t_0)$, then for every $x$,
the set of possible values of $\phi(x,t_0)$ is the entire real line. 
So, to get a {\it guaranteed} interval for $\phi(x,t_0)$ (and hence,
for the desired physical quantity), we need to use some additional
information. The process of using this additional information to get
non-trivial estimates for the solution of the inverse problem is
called a {\it regularization} \cite{TA77,I83,G84,I86,I86a,LRS86}.
There are several situations where this additional information is
available: 

\begin{itemize}

\item[$\bullet$]
If we are analyzing familiar processes, then
we usually know (more or less) how the desired function 
$\phi(x,t_0)$ looks like. For
example, we may know that $\phi(x,t_0)$ is a linear function
$C_1+C_2\cdot x_1$, or a sine function $C_1\cdot\sin(C_2x_1+C_3)$,
etc. In mathematical terms, we know that $\phi(x,t_0)=f(x,C_1,\ldots ,C_k)$,
where $f$ is a known expression, and the only problem is to
determine the coefficients $C_i$. This is how, for example, the
orbits of planets, satellites, comets, etc., are computed: 
the general shape of an orbit is known
from Newton's theory, so we only have to estimate the parameters of a
specific orbit. In such cases, the existence of several other
functions $\phi(x,t_0)$ 
that are consistent with the same observations, is not a big
problem, because we choose only the functions $x(t)$ that are
expressed by the formula $f(t,C_1,\ldots ,C_k)$.
This is not, however, a frequent situation in physics, because 
one of the main objectives (and the main
challenges) of physics is to analyze new phenomena, new effects, qualitatively
new processes, and in these cases no prior expression $f$ is known.

\item[$\bullet$]
In some cases, we
know the statistical characteristics of the reconstructed quantity 
$\phi(x,t_0)$ and statistical
characteristics of the
measurement errors. In these cases, we can formulate the problem of
choosing the maximally probable $\phi(x,t_0)$, and end up with one of the
methods of {\it statistical regularization}, or {\it filtering}
(Wiener filter is one of the examples of this approach). 

\item[$\bullet$]If we do not have this statistical information, but we know, 
e.g., that the average rate of change of $x(t)$ is
smaller than some constant $\Delta$ (i.e., $\sqrt{\int \dot
x(t)^2\,dt}\le\Delta$), then we can apply regularization methods
proposed by A. N. Tikhonov and others \cite{TA77,G84,LRS86}. 

\item[$\bullet$]
In many cases, we do not have the desired statistical information.
However, we may have some {\it expert knowledge}. 
For example, if we
want to know how the temperature on a planet changes with time $t$,
then the experts can tell that most likely, $x(t)$ is limited by
some value $M$, and that the rate $\dot x(t)$ with which the
temperature changes, is typically (or ``most likely,'', etc)
limited by some value $\Delta$, etc. We can also have some expert
knowledge about the error, with which we perform our measurements, so the
resulting expert's knowledge about the value of measured quantity $y$ 
looks like ``the difference between the measured value
$\tilde y$ and the actual value $y$ is most likely, not bigger than
$\delta$'' (where $\delta$ is a positive real number given by an
expert). 
The importance of this information is stressed in \cite{B92} and
in Chapter 5 of \cite{KM87}.
The methods of using this information and their application to testing
airplane and spaceship engines is described in \cite{KR86,K92,K92a,K92b}.
\item[$\bullet$] In many case, we do not have any quantitative expert
information like the one we described. In these cases,  
it is usually recommended to use
some heuristic (or semi-heuristic) regularization techniques 
\cite{TA77,I83,G84,I86,I86a,LRS86,CB86}. These methods often lead to
reasonable results, but they do not give any {\it guaranteed} estimate
for the reconstructed value $\phi(x,t_0)$. 
\end{itemize}
\item Suppose that we have the equations, and that we have chosen an
appropriate regularization for these equations. Then, in principle, we
can have the guaranteed estimate. The problem is that the numerical
methods that the physicists currently use do not give us these
guaranteed estimates. For example, we may have an iterative procedure
for solving the equation, and in this procedure, we stop if the next
iteration is close to the previous one. The fact that iterations are
close may mean that we are close to the actual solution, but how close
are we? In other words, how to get a {\it guaranteed estimate} for the
solution that is obtained by a heuristic method? For some equations,
such methods are known \cite{D86,DoS,Dobronets 1995}, but these
methods are far from being general.
\end{itemize}
There are several successful applications of interval methods (i.e.,
computational methods which provide guaranteed estimates) to
physics:
\begin{itemize}
\item {\it Stability of solar system and likewise systems}: 
there exists a K.A.M.
method (Kolmogorov-Arnold-Moser) that proves that for sufficiently
small values of some parameter, the Solar system is stable. The upper
bound is much lower than the actual value \cite{Moser 1978}. In 
\cite{Rana 1987,Llave 1987,Llave 1990,Llave 1990a,Llave 1991a}, 
interval computations are used to find upper
bounds for stability that for some systems, cover up to 90\% of the
actual stability zone. 
\item {\it Relativistic stability of matter} \cite{Fefferman
1986,Llave 1991}: for
relativistic version of Schroedinger equation, it is proved that for
sufficiently small charges, the
energy spectrum is non-negative (i.e., for $N\to\infty$, the system
does not collapse). The estimate for the charge is close to the one
for which the collapse actually occurs. 
\item {\it Asymptotic energy of atoms} 
\cite{Seco 1989,Seco 1991,Cordoba 1994} 
(based on Th. Fermi equation).
\end{itemize}
All these applications and the corresponding methods are
{\it domain-specific}. What can we do to get guaranteed 
estimates in the general case?

There are two possible approaches to this problem:
\begin{itemize}
\item A {\it pessimistic} approach: that we will never be able to get
guaranteed estimates. This approach is typical in {\it statistics}. For
example, a well-known statistician R. A. Fisher says that a
``hypothesis is never proved or established, but is possibly disproved,
in the course of experimentation'' (\cite{Fisher 1947}, p. 16; for a
modern description of this viewpoint, see, e.g., \cite{Howson 1989}).
Strictly speaking, from this viewpoint, 
we cannot even say that a theory is {\it 
disproved} with a guarantee. Indeed, if, e.g., a theory predicts 1, and the
measurement has led to 2, then, no matter how small the standard deviation
of the measurement error can be, the probability that the difference is
caused by the measurement error is non-zero, and so, it is possible
that the theory is still correct.
\item An {\it optimistic approach}, that most physicists hold, 
is that we {\it can} make guaranteed conclusions from the experiments.
A disproved theory {\it is} wrong, and the chance that the
measurement error has caused it is as large as having the cards in
order after thorough shuffling, or a possibility to win the lottery
every time by guessing the outcome: it is {\it impossible.} 
\end{itemize}
In this paper, we will describe a formalization of the optimistic
approach. 

\section{Main Idea in Detail: 
The Notion of ``Not Abnormal'' and How To Formalize It}

\subsection{Physicists Assume That Initial Conditions And Values Of
Parameters Are Not Abnormal}

To a mathematician, the main contents of a physical theory is the
equations. The fact that the theory is
formulated in terms of well-defined mathematical equations means that
the actual field must satisfy these equations. However, 
this fact does {\it not} mean that {\it every} solution of these
equations has a physical sense. Let us give two examples:
\begin{itemize}
\item At any temperature greater than absolute zero, particles are
randomly moving. It is theoretically possible that all the particles
start moving in one direction, and, as a result, the chair that I am
sitting on starts lifting up into the air. 
The probability of this event is small (but positive), so, from the purely
mathematical viewpoint, we can say that this event is possible but
highly unprobable. However, the physicists say plainly that such an
abnormal event is {\it impossible} (see, e.g., \cite{Feynman 1972}).
\item Another example from statistical physics: Suppose that we have a
two-chamber camera. The left chamber if empty, the right one has gas in
it. If we open the door between the chambers, then the gas would
spread evenly between the two chambers. It is theoretically possible
(under appropriately chosen initial conditions) that the gas that was
initially evenly distributed would concentrate in one camera, but
physicists believe this abnormal event to be impossible. This is a
general example of what physicists call {\it irreversible processes}:
on the atomic level, all equations are invariant with respect to
changing the order of time flow $t\to -t$). So, if we have a process
that goes from state $A$ to state $B$, then, if at $B$, we revert all
the velocities of all the atoms, we will get a process that goes from
$B$ to $A$. However, in real life, many processes are clearly
irreversible: an explosion can shatter a statue, but it is hard to
imagine an inverse process: an implosion that glues together shattered
pieces into a statue. Boltzmann himself, the 19 century author of 
statistical physics, explicitly stated that such inverse processes 
``may be regarded as impossible, even though from the viewpoint of
probability theory that outcome is only extremely improbable, not
impossible.'' \cite{Boltzmann 1877} (for this similar citations from other
founding fathers of statistical physics, see \cite{Kuhn 1978}).  
\item If we flip a coin 100 times in a row, and get heads
all the time, then a person who is knowledgeable in probability would
say that it is possible, while a physicist (and any person who uses 
common sense reasoning) would say that the coin is not fair, because
it is was a fair coin, then this abnormal event would be impossible. 
To illustrate this point, G. Polya in \cite{Polya 1968} cites
the following anecdote from the treatise of J. Bertrand on probabilities: 
\begin{itemize}
\item[] One day in Naples the reverend Galiani saw a man from the
Basilicate who, shaking three dice in a cup, wagered to three sixes;
and, in fact, he got three sixes right away. Such luck is possible, 
you say. Yet the man succeeded a second time, and the bet was repeated.
He put back the dice in the cup, three, four, five times, and each
time he produced three sixes. ``Sangue di Bacco'', exclaimed the
reverend, ``the dice are loaded!'' And they were.
\end{itemize}
\item In all the above cases, we knew something about probability.
However, there are examples of this type of reasoning in which
probability does not enter into picture at all. For example, in
general relativity, it is
known that for almost all initial conditions (in some reasonable
sense) the solution has a singularity point. Form this, physicists
conclude that the solution that corresponds to the geometry 
of the actual world has a singularity (see, e.g., \cite{Misner 1973}):
the reason is that the initial conditions that lead to a
non-singularity solution are abnormal (atypical), and the actual
initial conditions must be not abnormal. 
\end{itemize}
In all these cases, the physicists (implicitly or explicitly) require
that the actual values of the fields must not satisfy the equations,
but they must also satisfy the additional condition: that the initial
conditions should {\it not} be {\it abnormal}. 

\subsection{The Notion of ``Not Abnormal'' Is Difficult to Formalize}
At first glance, it looks like in the probabilistic case, this
property has a natural formalization: if a probability of an event is
small enough (say, $\le p_0$ for some very small $p_0$), then 
this event cannot happen. 
For example, the probability that a fair coin falls heads 100 times in
a row is $2^{-100}$, so, if we choose $p_0\ge
2^{-100}$, then we will be able to conclude that such an event is
impossible.
The problem with this approach is that {\it every} sequence of heads
an details has exactly the same probability. So, if we choose
$p_0\ge 2^{-100}$, we will thus exclude all possible sequences of heads
and tails as physically impossible. However, anyone can toss a coin 100
times, and this prove that some sequences are physically possible.
\smallskip

\noindent 
{\it Historical comment.} This problem was first noticed by Kyburg
under the name of {\it Lottery paradox} \cite{Kyburg 1961}: 
in a big (e.g., state-wide)
lottery, the probability of winning the Grand Prize 
is so small, then a reasonable person should not expect it. However,
some people do win big prizes (for a recent discussion of this
paradox, see, e.g., \cite{C89,Etherington 1990,Pollock 1990}). 

\subsection{How to Formalize The Notion of ``Not Abnormal'': Idea}
``Abnormal'' means something unusual, rarely happening: if something
is rare enough, it is not typical (``abnormal''). Let us
describe what, e.g., an abnormal height may mean. If a person's height
is $\ge 6$ ft, it is still normal (although it may be considered
abnormal in some parts of the world). Now, if instead of 6 pt, we
consider  6 ft 1 in, 6 ft 2 in, etc, then sooner or later we will end
up with a height $h$ such that everyone who is higher than $h$ will
be definitely called a person of abnormal height. We may not be sure
what exactly value $h$ experts will call ``abnormal'', but we are sure
that such a value exists. 

Let us express this idea is general terms. We have a {\it Universe of
discourse}, i.e., a set $U$ of all objects that we will consider.
Some of the elements of the set $U$ are abnormal (in some sense), and
some are not. Let us denote the set of all elements that are {\it
typical} ({\it not abnormal}) by $T$. On
this set, we have a decreasing sequence of sets $A_1\supseteq
A_2\supseteq \ldots \supseteq A_n\supseteq \ldots $ with the property that
$\cap A_n=\emptyset$. In the above example, $U$ is the set of all
people, $A_1$ is the set of all people whose height is $\ge 6$ ft,
$A_2$ is the set of all people whose height is $\ge 6$ ft 1 in, $A_2$
is the set of all people whose height is $\ge 6$ ft 2 in, etc. 
We know that if we take a sufficiently large $n$, then all elements of
$A_n$ are abnormal (i.e., none of them belongs to the set $T$ of not
abnormal elements). In mathematical terms, this means that for some
$n$, we have $A_n\cap T=\emptyset$.

In case of a coin: $U$ is the set of all infinite sequences of results
of flipping a coin; $A_n$ is the set of all sequences that start with
$n$ heads but have some tail afterwards. Here, $\cup A_n=\emptyset$.
Therefore, we can conclude that there exists an $n$ for which all
elements of $A_n$ are abnormal. So, if we assume that in our world,
only not abnormal initial conditions can happen, we can conclude that
for some $n$, the actual sequence of results of flipping a coin cannot
belong to $A_n$. The set $A_n$ consists of all elements that start
with $n$ heads and a tail after that. So, the fact that the actual
sequence does not belong to $A_n$ means that if an actual sequence has
$n$ heads, then it will consist of all heads. In plain words, if we
have flipped a coin $n$ times, and the results are $n$ heads, then
this coin is biased: it will always fall on heads. 

Let us describe this idea in mathematical terms.

\subsection{Formal Definition \cite{Finkelstein 1987}}

To make formal definitions, we must fix a formal theory: e.g., the set 
theory ZF (the definitions and results will not depend on what exactly
theory we choose). 
\smallskip

\noindent
{\bf Definition 1.} {\sl We say that a set $S$ is {\it definable} if 
in ZF, there exists a formula $P(x)$ with one free variable $x$ 
such that $P(x)$ iff $x\in S$.}
\smallskip

\noindent {\it Comment.}
Crudely speaking, a set is definable if we can {\it define} it in ZF. 
The set of all real numbers, the set of all solutions of a
well-defined equations, every set that we can describe in mathematical
terms is definable. This does not means, however, that {\it every} set
is definable: indeed, every definable set is uniquely determined by
formula $P(x)$, i.e., by a text in the
language of set theory. There are only denumerably many words and
therefore, there are only denumerably many definable sets. Since,
e.g., there are more than denumerably many set of integers, some of
them are this not definable. 
\smallskip

\noindent 
{\bf Definition 2.} {\sl We say that a sequence of sets 
$A_1,\ldots ,A_n,\ldots $ is {\it definable} if in ZF, there exists a formula
$P(n,x)$ such that $x\in A_n$ iff $P(n,x)$.}
\smallskip

\noindent {\bf Definition 3.} {\sl 
\begin{itemize}
\item Let a set $U$ be given. We will call it a
{\it universal set}. 
\item A non-empty set $T\subseteq U$ is called a {\it set of
typical (not abnormal) elements} if for every definable sequence of sets
$A_n$ for which $A_n\supseteq A_{n+1}$ and $\cap A_n=\emptyset$, there
exists an $N$ for which $A_N\cap T=\emptyset$. 
\item If $u\in T$, we will say
that $u$ is {\it not abnormal}. 
\item For every property $P$, we say that 
{\it ``normally, for all $u$, $P(u)$''} if $P(u)$ is true for all $u\in
T$.
\end{itemize}}

\subsection{Existence Theorems}

The trivial existence result is: for every set $U$, there is a 
set of typical elements $T$ that satisfies Definition 3: indeed, we can
take a one-element set $T=\{u\}\in U$.

A more interesting existence result appears if we
take into consideration the fact that 
our definition did not completely 
capture the following property of the notion of
``abnormal'': that exceptions (i.e., abnormal elements) should be
rare. ``Rare'' usually means that the probability of an element being
abnormal should be small enough (i.e., $\le\varepsilon$ for some
given $\varepsilon>0$). We may not know the exact
probabilities, so we may want to choose the set $T$ in such a way that
exceptions will be rare no matter what probability measure we choose. 
To describe this situation, we thus need to fix a real number
$\varepsilon>0$, and a finite sequence of probability measures
$p_1,\ldots ,p_n$. The only problem with this idea is that definable sets
may be not measurable. Therefore, in order to apply it, we will modify
Definition 3 so that it will allow only sequence $A_n$ whose elements
are measurable w.r.t. given measures.
\smallskip

\noindent {\bf Definition 3$'$.} {\sl 
\begin{itemize}
\item Let a set $U$ be given. We will call it a
{\it universal set}. Let $p_1,\ldots ,p_m$ be probability measures on 
$U$. 
\item A non-empty set $T\subseteq U$ is called a {\it set of
$(p_1,\ldots ,p_m)-$typical elements} if for every definable 
sequence of sets 
$A_n$ for which $A_n\supseteq A_{n+1}$, $\cap A_n=\emptyset$, and all
elements $A_n$ are measurable w.r.t. each measure $p_i$, there
exists an $N$ for which $A_N\cap T=\emptyset$. 
\end{itemize}}
\medskip

\noindent {\bf PROPOSITION 1.} {\it Assume that we have a set $U$, $m$
probability measures $p_1,\ldots ,p_m$ on $U$, and a real number
$\varepsilon>0$. Then, there exists a set $T$ of 
$(p_1,\ldots ,p_m)-$typical elements for
which $p_i(T)\ge 1-\varepsilon$ for all $i$.}
\medskip

\noindent{\it Comment.}
For reader's convenience, all the proofs are given in the special
section at the end of the paper. 
\smallskip

In other words, it is possible to define abnormal elements in such a
way that for each of $m$ measures, the probability of an element to be
abnormal is $\le\varepsilon$.

Now, that we have the definition, let us show that this notion can
indeed help to give guaranteed estimates. 

\section{Based On Finitely Many Experiments, We Can Guarantee That The
Theory Is Correct}

\subsection{General Result}

Let us show first that if we assume that the results of experiments
are required not to be abnormal, then we can (potentially) {\it
guarantee} that the theory is correct after only finitely many
experiments. 

From the viewpoint of an experimenter, a physical theory can be viewed
as a statement about the results of physical experiments. If we had
an infinite sequence of experimental results $r_1,\ldots ,r_n,\ldots $, 
then we will be able to
tell whether the theory is correct or not. So, a theory can be defined
as a set of sequences $r_1,\ldots ,r_n,\ldots $ that are consistent with its
equations, inequalities, etc. In real life, we only have
finitely many results $r_1,\ldots ,r_n$, so, we can only tell whether the
theory is {\it consistent} with these results or not, i.e., whether
there is an infinite sequence $r_1,\ldots ,r_n,\ldots $ that starts with the
given results that satisfies the theory. 

It is natural to require that
the theory be {\it physically meaningful} in the following sense: 
if all experiments confirm the theory, then this theory should be
correct. An example of a theory that is not physically meaningful is
easy to give: assume that a theory describes the results of tossing a
coin, and it predicts that at least once, there should be a tail. In
other words, this theory consists of all sequences that contain at
least one tail. Let us assume that actually, the coin is so biased
that we always have heads. Then, this infinite sequence does not
satisfy the given theory. However, for every $n$, the sequence of the
first $n$ results (i.e., the sequence of $n$ heads) is perfectly
consistent with the theory, because we can add a tail to it and get an
infinite sequence that belongs to the set $\cal T$. 

Let us describe this idea in formal terms.
\smallskip
\newpage

\noindent {\bf Definition 4.} {\sl 
\begin{itemize}
\item Let a definable 
set $R$ be given. Its elements will be
called {\it possible results of experiments.} By $S$, we 
will denote the set of all possible sequences $r_1,\ldots ,r_n,\ldots $, each
element $r_i$ of which is a result of an experiment (i.e., $r_i\in R$). 
\item By a {\it theory}, we mean
a definable 
subset $\cal T$ of the set of all infinite sequences $S$. If $r\in
{\cal T}$, we say that an infinite sequence $r$ {\it satisfies} the
theory $\cal T$, or, that for this sequence $r$, {\it the theory $\cal
T$ is correct}. 
\item We say
that a finite sequence $(r_1,\ldots ,r_n)$ is {\it consistent} with the
theory $\cal T$ if there exists an infinite sequence $r\in {\cal T}$
that starts with $r_1,\ldots .,r_n$ and that satisfies the theory. In this
case, we will also say that the {\it first $n$ experiments confirm the
theory}.
\item We say that a theory $\cal T$ is {\it physically meaningful} if
the following is true: 
\begin{itemize}
\item[] Let $r$ be a sequence $r\in S$ such that for every $n$, the
results of first $n$ experiments from $r$ conform the theory $\cal T$.
Then, the theory $\cal T$ is correct for $r$. 
\end{itemize}
\end{itemize}}

In this case, the universal set consists of all possible infinite
sequence of experimental results, i.e., $U=S$. 
\medskip

\noindent {\bf PROPOSITION 2.} {\it For every physically meaningful 
theory $\cal T$, there exists an
integer $N$ such that normally, if the first $N$ experiment confirm
the theory $\cal T$, then this theory $\cal T$ is correct.}
\medskip

In other words, if the sequence of results $r$ is not abnormal, and
the results of first $N$ experiments are consistent with the theory,
then the theory is correct. This result shows that we can {\it
confirm} the theory based on finitely many observations. 
\smallskip

\noindent 
{\it Philosophical comment: physical induction and its paradoxes.}
The derivation of a general theory from finitely many experiments is
called {\it physical induction} (as opposed to {\it mathematical
induction}). These have been many formalizations of different ideas
that physicists use, and these formalizations has lead to successful
programs that can find a general dependency from the cases (see, e.g.,
\cite{Barzdin 1991}. However, in spite of the success in describing
several underlying ideas, the general 
physical induction is difficult to formalize, to the
extent that  a prominent philosopher 
C. D. Broad has called the unsolved problems concerning
induction a {\it scandal of philosophy} \cite{Broad 1952}. We can say that 
the notion of ``not abnormal'' justifies physical induction (and thus
resolves the corresponding scandal).  
\smallskip

\noindent{\it Philosophical comment: Ockham's Razor justified.}
Ockham Razor is a principle according to which, one should not
unnecessarily multiply the number of entities. It is usually
understood as follows: if we have two properties $A$ and $B$, and if in all
experiments, these two properties coincide, then we should assume that
these two properties are identical. The above result justifies
Ockham's razor principle: namely, if, as a theory $\cal T$, we
consider a statement that $A$ and $B$ are always identical, 
then from Proposition 2, we can conclude that normally, if $A$ and $B$
are identical for first $N$ experimental results, then these two
properties do always coincide. 

\subsubsection{Abnormal Theories and Related Paradoxes}
The necessary number of experiments differ from a theory to a theory.
In principle, we can formulate a theory that predicts the same results
as our normal physics until some arbitrarily chosen 
year $Y$, and then predicts something
else. For this theory, $N$ must be so large as to stretch to that year
$Y$. So, the larger $Y$, the larger $N$. Such theories are artificial
and abnormal. It turns out that if we restrict ourselves by {\it not
abnormal} theories, then we will have a universal bound on $N$:

Namely, let us assume that on the set $U$ of all pairs $({\cal T},r)$,
where $\cal T$ is a physically meaningful theory, and $r$ is a
sequence of experimental results, we have 
selected a subset of typical (not abnormal) pairs $T$.
Then, the following result is true:
\medskip

\noindent 
{\bf PROPOSITION 3.} {\it There exists an integer $N$ such that normally, if
a physically meaningful 
theory $\cal T$ is confirmed by the first $N$ experiments, then this
theory is correct.}
\medskip

In other words, if a pair $({\cal T},r)$ is not abnormal, and first $N$
experiments from $r$ confirm the theory $\cal T$, then $\cal T$ is
correct on $r$. 
\smallskip

\noindent
{\it Philosophical comment: Goodman's paradox explained.}
Formalization of physical induction is a difficult task, known to lead
to paradoxes. For example, Nelson Goodman 
\cite{Goodman 1946,Goodman 1955} has proposed the following paradox.
We have observed emeralds many times, we know that they are green, so
we conclude that emeralds are always green. Instead of the theory
``emeralds are green'', let us consider an alternative theory
``emeralds are grue'', where {\it grue} stands for ``green before the
year 2010, and blue after the year 2010''. Then, all the evidence that
we have used to conclude that emeralds are green, also confirms that
they are grue. However, to conclude that emeralds are grue is strange.

There have been several idea on how to solve this paradox (see, e.g., 
\cite{Barker 1960,Salmon 1963,Smokler 1966,F67,H69,B73,C89}). 
However, as it is remarked in Chapter 14 of 
\cite{Kyburg 1968}, these ideas has not yet lead to a consistent
formalization, and hence, Goodman himself did not
consider this paradox solved. 
From the physicist's viewpoint (that
Goodman himself explained in his papers and that other researchers
tried to formalize), 
Goodman's paradox is not a paradox at all: green is a natural
property, while grue is an abnormal one. The problem is to formalize
this distinction. The above formalization provides exactly this answer to
the paradox: 
Indeed, if a not abnormal theory $\cal T$ is confirmed by $N$ experiments and
is, therefore, correct, and if another theory $\cal T'$ is confirmed
by these same experiments, but leads to different predictions, then
due to Proposition 3, it means that $\cal T'$ is an abnormal theory.

Of course, this is not a complete solution of the related set of
problems, because we still need to find out how to distinguish between
normal and abnormal theories. 

\subsection{How To Guarantee The Exact Values of Parameters}

In the following text, by $\|x\|$, we will mean a Euclidean norm of
the vector $x$. 
\smallskip

\noindent{\bf PROPOSITION 4.} {\it Let $d$ be an integer, and let $a$ be a
definable point from $R^d$. Then, there exists an $\varepsilon>0$ such
that if $x$ is not abnormal, and $\|x-a\|\le \varepsilon$, then $x=a$.}
\medskip

\noindent{\it Comment.} 
This result is actually correct for an
arbitrary definable metric space $X$ with a metric $d$. 
\smallskip

This means that for every set of typical elements $T\subseteq R^d$,
there exists an $\varepsilon>0$ such that if $x$ is not abnormal (i.e.,
$x\in T$), and $x$ is $\varepsilon-$close to $a$, then $x=a$.

In other words, if we make an assumption (that physicists usually
make) that the actual values of parameters are not abnormal, then 
it is not true that we can test a theory with better and better
accuracy and never be 100\% guaranteed that the theory is correct:
there exists an $\varepsilon>0$ such that if we have confirmed the theory
with the accuracy $\varepsilon$, then this theory is true. 

From this result, we can conclude that {\it coincidences are not
accidental}: Indeed, suppose that we know a constant $c$. Now, if in
some other (not clearly related) physical experiments we get a
constant that is very close to $c$, then we can conclude that this is
the same constant. This type of argument is very frequent in physics:
e.g., the discovery that light consists of electromagnetic waves was 
prompted by the fact that the computed velocity of these waves turned
out to be very close to the measured speed of light. 
\smallskip

\noindent
{\it Comment.} Arguments of the type {\it ``This is too improbable to be a
mere coincidence. There must be some reason.''} are also used in {\it
mathematics}, to deduce hypotheses from the observed facts (see, e.g.,
\cite{Polya 1968}, Chapter XIV, Section 16). 

\section{Restriction To ``Not Abnormal'' Solutions Leads To
Regularization Of Ill-Posed Problems}

The material described in this section follows \cite{Kozlenko 1984}. 

\subsection{The Main Result}

An ill-posed problem arises when we want to reconstruct the state $s$
from the measurement results $r$. Usually, all physical dependencies
are continuous, so, small changes of the state $s$ result in small
changes in $r$. In other words, a mapping $f: S\to R$ from the set of
all states to the set of all observations is continuous (in some
natural topology). We consider the case when the measurement results
are (in principle) sufficient to reconstruct $s$, i.e., the case when
the mapping $f$ is 1-1. That the problem is ill-posed means that 
small changes in $r$ can lead to huge changes in $s$, i.e., that the
inverse mapping $f^{-1}:R\to S$ is {\it not} continuous. 

We will show that if we restrict ourselves to states $S$ that are not
abnormal, then the restriction of $f^{-1}$ will be continuous, and the
problem will become well-posed.
\smallskip

\noindent
{\bf Definition 5.} {\sl A definable metric space $(X,d)$ is called {\it
definably separable} if there exists a definable everywhere
dense sequence $x_n\in X$.}
\smallskip

\noindent
{\bf PROPOSITION 5.} {\it Let $S$ be a definably separable definable
metric space, $T$ be a set of all not abnormal elements of $S$, 
and $f:S\to R$ be a continuous 1-1 function. Then, 
the inverse mapping $f^{-1}:R\to S$ is continuous for every $r\in
f(T)$.}
\smallskip

In other words, if we know that we have observed a not abnormal state
(i.e., that $r=f(s)$ for some $s\in T$), then the reconstruction
problem becomes well-posed. So, if the observations are accurate
enough, we get as small guaranteed intervals for the reconstructed
state $s$ as we want. 
\smallskip

\noindent{\it Mathematical comment.} 
This Proposition uses the following Lemma that may be of independent
interest:
\medskip

\noindent{\bf LEMMA 1.} {\it If $X$ is a definably separable definable
metric space, and $T$ is a set of all not abnormal elements of $X$, 
then the closure $\overline T$ is a compact set.}

\subsection{How Can We Actually Use This Result to Get Guaranteed
Estimates \cite{Kozlenko 1984}}

To actually use this result, we need an {\it expert} who will tell us
what is abnormal. We will show that if we use such an expert, then for
every computable function $f:X\to Y$, if we know that $x\in T$, then
sufficiently accurate knowledge of $f(x)$ will enable us to
reconstruct $x$ with any given accuracy.
\smallskip
\newpage

\noindent {\bf Definition 6.} {\sl 
\begin{itemize}
\item By an {\it expert},
we mean a mapping $E:\{A_n\}\to Z$ that transforms a definable
decreasing sequence with an empty intersection into an integer $N$
for which $A_N\cap T=\emptyset$ (i.e., for which all elements from
$A_N$ are abnormal).
\item We say that an output is 
{\it computable with an expert}
if it is computable on a computer that can consult an expert (i.e.,
that sends an expert a formula defining $\{A_n\}$ and gets $N$).
\end{itemize}}
\smallskip

The following definitions are standard in constructive analysis 
\cite{B67,B79,BB85,Nakamura 1993}.
\smallskip

\noindent {\bf Definition 7.} {\sl 
\begin{itemize}
\item We say that 
an algorithm $\cal U$ {\it computes} a real number $x$ if
for every natural number $k$, it generates a rational number $r_k$
such that $|r_k - x| \le 2^{-k}$. We say that we have a {\it
computable} real number if we have an algorithm $\cal U$ that computes
it.
\item By a {\it computable separable metric
space}, we understand a separable metric space $(X,d)$ with an
everywhere dense sequence $\{x_n\}$ for which there exists an
algorithm, that transforms a pair of
positive integers $n,m$ into a computable real number
$d(x_n,x_m)$.
\item By a {\it computable element} of a computable space we
understand a pair consisting of an element $x\in X$ and an algorithm
that given $n$, returns an integer $m(n)$ for which 
$x_{m(n)}$ is a $2^{-n}-$approximation to $x$. 
\item 
Let $X$ and $Y$ be computable separable metric spaces. 
We say that an algorithm $\cal V$ {\it
computes} a function $f:X\to Y$ if $\cal V$ includes calls to an
(unspecified) algorithm $\cal U$ so that when we take as $\cal U$ an
algorithm that computes an element $x\in X$, then $\cal V$ will
compute an element $f(x)\in Y$.
\item We say that a computable function $f$ is
{\it constructively continuous} on a set $S$ if there exists an
algorithm, that for every $\varepsilon>0$, generates $\delta>0$ such
that if $|x-y|\le\delta$, then $|f(x)-f(y)|\le \varepsilon$. 
\end{itemize}}
\smallskip
\newpage

\noindent 
{\bf Definition 8.} {\sl Assume that we are given the following
information:
\begin{itemize}
\item computable separable metric spaces $X$ and $Y$;
\item a computably continuous 
computable 1-1 function $f:X\to Y$ (1-1 means that if $x\ne x'$, then
$f(x)\ne f(x')$);
\item an algorithm that, given an integer $k$, returns 
a $2^{-k}-$approximation to $f(x)$, where $x\in T$ is an (unknown)
typical element;
\item a positive integer $l$.
\end{itemize}
We say that algorithm {\it solves the inverse problem} if, given the
above information, this algorithm returns a $2^{-l}-$approximation to
$y$. If such an algorithm exists, then we will say that an inverse
problem is {\it computable}.}
\medskip

\noindent 
{\bf PROPOSITION 6.} {\it The inverse problem is computable with an
expert.} 
\medskip

\noindent{\it Comment.} 
\begin{itemize}
\item This general algorithm can be applied to different numerical
problems: to solving a system of non-linear equations (when $X$ and $Y$
are $R^k$ for some $k$), to solving integral equations (when $X$ and
$Y$ are sets of functions), etc.
\item If we do not restrict ourselves to not abnormal elements $x$,
then in many cases, it will be impossible to have an algorithm for
solving the inverse problem: indeed, if such an algorithm is possible,
then the inverse function $f^{-1}$ is continuous, but, as we have
already mentioned, for some
continuous 1-1 mappings $f$ from a non-compact set, the inverse is not
continuous \cite{TA77}. 
\item The algorithm described in the proof is general and therefore
(as many general algorithms), when applied to simple problems, 
it may require unnecessarily many computation steps. 
There are cases when simpler methods are possible: e.g., if the signal
that we are trying to reconstruct is a smooth function, then we can 
ask an expert what is the upper bound for 
the signal's energy $\int (x'(t))^2\, dt$, and 
then use known regularization techniques \cite{TA77}.
\end{itemize}

\section{If We Impose The Condition That The Actual State Is Not
Abnormal, Then We Can Get Guaranteed Estimates Even For Heuristic
Numerical Methods}

\subsection{How To Check Whether A Numerical Method Always Works}

To check the numerical method, we can run it on several tests. 
Usually, the first tests are are simple computer-generated ones. If a
method behaves nicely on these simple tests, then it is tried on
realistic or real-life examples, where the input data come from real
experiments. So, we have a (potentially infinite) sequence of tests. 
Let us assume that testing is performed in such a way that on some
stage, every part and every aspect of the method is tested.
Mathematically, let us assume that the potentially infinite sequence 
of test cases has the following {\it completeness} property: 
if a method works correctly for all (infinitely many) 
test cases, then this method is always correct. 
\smallskip

\noindent{\bf Definition 9.} {\sl 
\begin{itemize}
\item Let a definable 
set $t$ be given. Its elements will be
called {\it tests.} By a {\it testing method} $S$, we mean
a set of infinite sequences $t_1,\ldots ,t_n,\ldots $ of tests.
\item By a {\it numerical method}, we mean
a definable 
subset $M$ of the set of all tests $t$. If $t_i\in M$, we say that a
method $M$ {\it passed} the test $t_i$; else, that the method $M$ {\it
failed} the test $M$. 
\item We say that a method is {\it correct} if it passes all tests
from $t$.
\item Let a class of methods $\cal M$ be fixed. 
We say that a testing method $S$ is {\it complete} for methods from
the class $\cal M$ if for every
sequence $(t_1,\ldots ,t_n,\ldots )\in S$, and for every method $M\in {\cal
M}$, if the method $M$ passes all the tests $t_1,\ldots ,t_n,\ldots $, then
this method is correct.
\end{itemize}}
\medskip

\noindent 
{\bf PROPOSITION 7.} {\it Let a testing method $S$ be complete for a
class of numerical methods $\cal M$. Then, for every method $M\in
{\cal M}$, there exists an integer $N$ such that if a sequence
$(t_1,\ldots ,t_n,\ldots )\in S$ is not abnormal, and 
$M$ passes the first $N$ tests of this
sequence, then this method $M$ is correct.}
\medskip

\noindent
This Proposition justifies the usual testing of a method, in which we
make a conclusion about its correctness after only finitely many
tests. The crucial assumption here is that we assume that the testing
sequence is taken from the real-life examples, and these examples are
not abnormal. 

\subsection{How To Get A Guaranteed Estimate For The Result}

In many practical cases, we know the process $x_k$ that is proven to
converge to the desired solution $x$, but we do not know when to stop
in order to guarantee the given accuracy $\varepsilon$ (i.e., to
guarantee that $d(x,x_k)\le\varepsilon$). 
\begin{itemize}
\item[]For example, we may use an
iterative method $x_{k+1}=F(x_k)$ to solve the equation $F(x)=x$. 
\end{itemize}
In these cases, heuristic
methods are used. There are two main groups of heuristic methods:
\begin{itemize}
\item Usually, in iterative methods, if $x_k=x_{k+1}$, then $x_n$ is
the required solution. Therefore, if $x_k$ and $x_{k+1}$ are close, we
can conclude that we are {\it close} to the solution. Hence,
we stop when the consequent values $x_k$ become close enough,
i.e., when $d(x_k,x_{k+1})\le\delta$ for some $\delta>0$. This method
is often used in physics, if, e.g., we have the expression of $x$ as a
sum of the infinite series (e.g., Taylor series in perturbation
methods). Then, if, e.g., second order terms are negligibly small, we
neglect quadratic {\it and} higher order terms, and use the linear
expression as an approximation to the desired solution (see, e.g.,
\cite{F65}). 
\item If we are solving the equation $f(x)=y$, then we stop when
$f(x_k)$ becomes small enough (i.e., when $d(f(x),y)\le\delta$ for
some $\delta>0$).
\end{itemize}
These stopping criteria can be described by the following general
definition:
\smallskip

\noindent
{\bf Definition 10.} {\sl Let $X$ be a definable metric space, and let
$S$ be a definable set of convergent sequences of $X$. 
\begin{itemize}
\item Let $\{x_k\}\in S$, $k$ be an integer, and $\varepsilon>0$ a
real number. We say that $x_k$
is {\it $\varepsilon-$accurate} if $d(x_k,\lim x_p)\le\varepsilon$.
\item Let $d\ge 1$ be
an integer. By a {\it
stopping criterion}, we mean a function $c:X^d\to R^+_0=\{x\in
R\,|\,x\ge 0\}$ that satisfies
the following two properties:
\begin{itemize}
\item[$\bullet$] If $\{x_k\}\in S$, then $c(x_k,\ldots ,x_{k+d-1})\to 0$.
\item[$\bullet$] If for some $\{x_k\}\in S$ and for some $n$,
$c(x_k,\ldots ,x_{k+d-1})=0$, then $x_k=\ldots =x_{k+d-1}=\lim x_p$.
\end{itemize}
\end{itemize}}

\noindent
The two above-described criteria correspond to $c(x,x')=d(x,x')$ and
$c(x)=d(f(x),y)$.
\medskip

\noindent 
{\bf PROPOSITION 8.} {\it Let $c$ be a stopping criterion. Then, for
every $\varepsilon$, there exists a $\delta>0$ such that if a sequence
$\{x_k\}$ is not abnormal, and $c(x_k,\ldots ,x_{k+d-1})\le\delta$, then
$x_k$ is $\varepsilon-$accurate.}
\medskip

\noindent 
So, if we restrict ourselves to not abnormal sequences only (i.e.,
sequence that stem from not abnormal, physical observations), then
$c(x_k,\ldots ,x_{k+d-1})\le\delta$ guarantees that we are
$\varepsilon-$close to the desired solution. In particular,
$d(x_k,x_{k+1})\le\delta$ and $d(f(x_k),y)\le\delta$ guarantee that
$d(x_n,x)\le\varepsilon$. In case we are summing a numerical series
$x_k=a_1+\ldots +a_k$, we have $d(x_k,x_{k+1})=|a_{k+1}|$, so, this
stopping criterion means that means if the next term is negligible
($|a_{k+1}|\le\delta$), then we are $\varepsilon-$close to the sum:
$|x_k-x|\le\varepsilon$.

\subsection{When Will The Algorithm Stop?} 

In practice, it is not sufficient to claim that an algorithm generates
a guaranteed estimate. We would like to know when to expect the
result. A computer can go wrong, so if the computations take too
long, we would like to know whether it is just taking long, or there
has been a computer error, and we better start anew. 

Theoretically, arbitrarily long computations are possible. However,
as we will see, computations do not take too long. 
\smallskip

\noindent{\bf Definition 11.} {\sl 
\begin{itemize}
\item Let a set $U$ be given. Its elements are
called {\it computations.} 
\item Let a function $t:U\to R\cup\{\infty\}$ be
given. The value $t(u)$ will be called the {\it run time} of the
computation $u$. 
\item We say that computation {\it terminates} if $t(u)\ne\infty$.
\end{itemize}} 
\medskip

\noindent
{\bf PROPOSITION 9.} {\it There exists a number $T_0>0$ such that if a
computation $u$ is not abnormal, and it terminates, then its run time 
is $\le T_0$.}

\subsection{If a Numerical Method is Polynomial-Time and Not
Abnormal, Then It Is Truly Feasible}

It is well known that not all algorithms are realistic (see, e.g.,
\cite{Lewis 1981}, Section 7.1). If an
algorithm requires, say, $2^{2^{{\rm len}(x)}}$ 
computational steps for an input $x$ of
length ${\rm len}(x)$, 
then for realistic lengths (e.g., for ${\rm len}(x)=100$) this number of
steps will exceed the lifetime of the Universe (according to modern
cosmology). So if we are interested in separating purely theoretical
algorithms from the ones that can be actually run on the computers
(existing or future ones), we must somehow formalize the notion of
feasibility. 

The most widely used formalization of this notion is that feasible
algorithms are exactly the ones that are time-polynomial, i.e., the
ones for
which the running time is limited by some polynomial $P({\rm len}(x))$ of the
input length ${\rm len}(x)$ (see, e.g., \cite{Lewis 1981}, Section 7.4;
\cite{M91}, Ch. 23). 
There exist formal systems of reasonable axioms that justify
this choice (see, e.g., \cite{S80}). 

However, the majority of the 
researchers agree that this is not the precise description of a
feasible algorithm, because some time-polynomial algorithms are
evidently not feasible. For example, an algorithm that takes 
$10^{10^{10}}{\rm len}(x)$
time to compute is time-polynomial (even linear-time)
but it can hardly be called feasible: even for ${\rm len}(x)=1$,
it requires the
computation time that is exponentially bigger than the lifetime of
the Universe. 

There are two possible approaches to this situation:
\begin{itemize}
\item We can view this situation as a problem, and try to come up with
a new definition of feasibility that will really describe only
physically feasible algorithms. Such a formalization is proposed, e.g., in
\cite{Nguyen 1995}.  
\item In this section, we will pursue another approach: we will show
that normally, the values of the coefficient and the exponent cannot
grow indefinitely: namely, 
there exist $C$ and $K$ such that 
if we exclude abnormal methods, then running time $t_{\cal U}(x)$ 
for {\it all} not abnormal 
time-polynomial algorithms $\cal U$ is limited by $C\cdot ({\rm len}(x))^K$.
\end{itemize}
\medskip

\noindent
{\bf PROPOSITION 10.} {\it There exists $C>0$ and $K>0$ such that
if a polynomial-time algorithm $\cal U$ is not abnormal, then its
running time $t_{\cal U}(n)$ is bounded by $C\cdot n^K$.}
\medskip

\noindent
In other words, if we take the set of all polynomial-time algorithms
as $U$, and denote by $T$ the set of all typical elements of $U$, then
there exists $K$ and $C$ such that the running time of every algorithm
from $T$ is bounded by $C\cdot n^K$.

This proposition can be confirmed by the fact that for every problem,
for which a time-polynomial algorithm has been known for some time
(for a few years), in a few years, a new algorithm is discovered for
which the running time is limited by a cube of $n$, i.e., by 
$C\cdot n^3$ (see, e.g.,
\cite{A}; the time during which this happens is jokingly called
``{\it incubation} period''). 

\section{The Notion Of ``Not Abnormal'' Is Also Helpful For 
Foundations of Physics}

In the previous sections, we described how the notion of ``not
abnormal'' can lead us to guaranteed interval (and, in general, error)
estimates. In other words, we showed that this notion is helpful in
{\it computational physics}. 
In this section, we will show that this notion can also
help in the problems related to {\it foundations of physics}. 
These results will show that our formalization is in good accordance 
with the modern theoretical physics. 

Some of these results have been presented in 
\cite{Kosheleva 1978,Finkelstein 1987}.

\subsection{Every Physical Quantity is Bounded}

\noindent
{\bf PROPOSITION 11.} {\it If $U$ is a definable set, and $f:U\to R$ is
a definable function, then there exists a number $C$ such that if
$u\in U$ is not abnormal, then $|f(u)|\le C$.} 
\medskip

\noindent
If we use the physicists' idea that
abnormal initial conditions and/or abnormal values of parameters are
impossible, then we can make the following conclusions:
\smallskip

\noindent
{\bf Special relativity.} If as $U$, we take the set of all the
particles, and as $f$, we take velocity, then we can conclude
that the velocities of all (not abnormal) particles is bounded by some
constant $C$. This is exactly what special
relativity says, with the speed of light as $C$. 
\medskip

\noindent
{\bf Cosmology.} If we take the same state $U$, and as $f$, take the
distance from the a particle $u$ to some fixed point in the Universe,
then we can conclude that the distances between 
particles in the Universe are bounded by a constant $C$. 
So, the Universe is {\it finite}. Similarly, if we take a time
interval between the events as $f$, we can conclude that the Universe
has a {\it finite lifetime.}
\medskip

\noindent
{\bf Why particles with large masses do not exist.}
If we take mass of the particle as $f$, then we can conclude that the
masses of all particles are bounded by some constant $C$. This
result explains the following problem:
\begin{itemize}
\item 
Several existing particle classification schemes allow 
particles with arbitrarily large masses 
\cite{Griffiths 1987,Brink 1988,Nachtmann 1990}. 
E.g., in Regge
trajectory scheme, particles form families with masses $m_n=m_0+n\cdot
d$ for some constants $m_0$ and $d$. When $n\to\infty$, we have
$m_n\to\infty$.  
\item Only particles with
relatively small masses have been experimentally observed (see, e.g.,
\cite{PDL}). 
\end{itemize}
These particles with large masses, that are difficult to wed out using
equations only, can be easily weeded out if use the notion of ``not
abnormal''. 
\medskip

\noindent
{\bf Dimensionless constants are usually small.} This is the reason
why physicists can safely estimate and neglect, e.g., quadratic (or,
in general, higher order terms) in asymptotic expansions, even though
no accurate estimates on the coefficients on these terms is known
\cite{Landau 1951,L59a,L59,Landau 1960,Landau 1965,F65,Landau 1969}. In
particular, such
methods are used in quantum field theory, where we add up several
first Feynman diagrams \cite{F65,Berestetsky 1974,Nachtmann 1990}; 
in celestial mechanics 
\cite{Chebotarev 1967,Misner 1973,Taff 1985,Brumberg 1991}, etc. 
\smallskip

\noindent
{\it Comment: Consequences for philosophy of mathematics.} Physically
meaningful numbers are bounded. Hence, it seems reasonable to 
place only
physically meaningful integers in the 
foundations of mathematics. In the corresponding formalisms, there
will be finitely only finitely many integers. The ideas of such
formalisms were originally developed by Van Dantzig,
Essenine-Volpine, and Kolmogorov \cite{Kolmogorov 1968}, and 
have been later transformed into a useful formalism
by Parikh (see \cite{P71} and references therein). 
\newpage

\subsection{Quantization}

\noindent
{\bf PROPOSITION 12.} {\it Let $U$ be a definable set, and $f:U\to R$ be
a definable function. Then, there exists a number $\varepsilon>0$ such
that if $u$ is not abnormal and $f(u)\ne 0$, then
$|f(u)|\ge\varepsilon$.}
\medskip

\noindent
Together with the physicists' idea that abnormal situations are
impossible, we can conclude that
a physical quantity cannot have arbitrarily small
positive values: there must be the smallest value that is indivisible.
This explains, e.g., why the electric charge cannot tale any value we
want: there is the charge quantum ($1/3$ of an electron's charge). 
\medskip

\noindent{\bf Infinities in quantum field theory disappear.} 
This result can be justified by the joint use of 
quantization and boundedness: indeed, in quantum field theory,
infinities are caused by the fact that we have have to integrate over
all momenta $p$ of all particles. Infinities happen because we have to
integrate over $p\to\infty$ and $p\to 0$ (see, e.g., \cite{F65}). 
If we apply boundedness and
quantization results to momentum, we conclude that $p$ is bounded from
above and from below. Therefore, all the integrals should be finite
(another interval-related argument that makes infinities disappear is
given in \cite{Korlyukov 1995}). 
\medskip

\noindent{\bf Schroedinger's cat paradox stops being paradoxical}.
According to traditional quantum mechanics, states are described by a
vectors from a Hilbert space $L^2$, and all vectors have a physical
meaning. E. Schroedinger has shown that this assumption leads to the
following paradox (for description and discussions, see, e.g.,
\cite{Penrose 1989}): Suppose that we place a cat into a box with a
gun aimed at it. A gun is controlled by the switch, which can be
triggered by a left-polarized photon. If we send a photon in a
left-polarized state $s_1$, the gun fires, and the cat is dead. 
If we send a photon in a right-polarized state $s_2$, the gun does not
fire, and the cat is alive. Suppose now that we send a photon in a state
$s$ that is a superposition of $s_1$ and $s_2$ (i.e., in mathematical
terms, a linear combination). Equations of quantum mechanics are
linear, so, as a result, we get the state that is a superposition of
dead and alive. Such a superposition is difficult to imagine, because
in real life, an animal is either dead, or alive. 

This paradox is based on the assumption that {\it all} vectors from a Hilbert
space are physically meaningful states. If we impose the additional
condition that only not abnormal states are physically possible, then
we can exclude some states as being abnormal. 
Indeed, from the quantization result, it follows that there exists an
$\varepsilon$ such that if a physically meaningful 
state is $\varepsilon-$close to the
``dead'' state, then it {\it is} the ``dead'' state. Paradoxical
continuous transition between dead and alive thus disappears.

\subsection{Chaos}

\noindent{\bf The origin of chaos.} 
Restriction to not abnormal also explains the origin of
chaotic behavior of physical systems. In mathematical terms, chaos
means that after some time, the states of the system form a so-called
{\it strange attractor}, i.e., in topological terms, a {\it completely
disconnected set} in the following precise sense:
\smallskip

\noindent
{\bf Definition 12.} {\sl A set $S$ in a metric space $X$ is called {\it
completely disconnected} if for every $s_1,s_2\in S$, there exist open sets
$S_1$ and $S_2$ such that $s_1\in S_1$, $s_2\in S_2$, $S_1\cap
S_2=\emptyset$, and $S\subseteq S_1\cup S_2$.}
\smallskip

\noindent
In other words, every two points belong to different topological
components of the set $S$. The relationship between this definition and
typical elements is given by the following result:
\medskip

\noindent
{\bf PROPOSITION 13.} {\it In a definable separable metric space, the set
of typical elements is completely disconnected.}
\medskip

\noindent
So, if we assume (as physicists do) that abnormal states are impossible,
then we immediate arrive at the chaotic dynamics.
\medskip

\noindent
{\bf Spontaneous symmetry violations.} Equations of physics have lots
of symmetries. If an equation is, e.g., invariant w.r.t. rotations,
and the initial condition is rotation-invariant, then the solution
stays rotation-invariant for all moments of time. From the
mathematical viewpoint, symmetric solutions are quite possible.
However, in real life, we only observe {\it approximately symmetric
solutions}. E.g., in cosmology, from the observations of the 3K
radiation, we can conclude that the initial state of the Universe was
highly isotropic and homogeneous (see, e.g., \cite{Misner 1973}). However, 
the observed Universe is not. This means that the initial conditions
were only {\it approximately} isotropic and homogeneous. 

In each particular case, we may have specific physical 
reasons for symmetry violation. Restriction to ``not abnormal'' leads
to a general explanation \cite{Finkelstein 1987}: namely, 
with this restriction, the theory consists not only of the set if
equations, but also of the set $T$ of physically possible (not
abnormal) initial conditions. We are going to show that even if the
equations are invariant, the set $T$ is not. 
\smallskip

\noindent{\bf Definition 13.} {\sl Let $X$ be a topological space. 
\begin{itemize}
\item By a {\it
continuous transformation group}, we mean a connected continuous group
$G$ with a continuous mapping $a:G\times X\to X$ such that
$a_{g_1g_2}(x)=a_{g_1}(a_{g_2}(x))$.
\item A set $S$ is called {\it invariant} w.r.t. $G$ if $a_{g}(S)=S$
for all $g\in G$ (where $a_g(S)=\{a_g(s)\,|\,s\in S\}$). 
\item We say that the continuous transformation
group $G$ is {\it non-trivial} if $a_g(x)\ne x$ for some $g\in G$ and
for some not abnormal $x\in X$.
\end{itemize}}
\medskip

\noindent
{\bf PROPOSITION 14.} {\it Let $X$ be a definable separable metric
space, $T$ be the set of typical elements of $X$, and let $G$ be a
non-trivial continuous transformation group. Then, the set $T$ is not 
invariant w.r.t. $G$.}

\subsection{All The Processes In The World Are Connected
\cite{Finkelstein 1987}}
 
Let us assume that we are studying two physically unrelated 
processes. Let $X_1$
denote the set of all possible states (initial conditions) of the
first process, and $X_2$ denote the set of all initial conditions of
the second process. Then, for each process, we can formalize the
physicists' idea that initial conditions cannot be abnormal by saying
that $x_1\in T_1$ and $x_2\in T_2$, where $T_i\subseteq X_i$ are the
corresponding sets of typical elements. 

We could also analyze the two processes as a whole. The state of a
pair of processes can be characterized by a pair of states
$(x_1,x_2)$. Since the
processes are unrelated, the state $x_1$ of the first process cannot
influence the state of the second one. So, for every $x_1$, the set of
all possible states $x_2$ of the second process is $X_2$. Therefore,
the set of all possible pairs $(x_1,x_2)$ is equal to the set
$X=X_1\times X_2$ of all the pairs $(x_1,x_2)$, $x_i\in X_i$. We can now
formulate our ``not abnormal'' idea by saying that $(x_1,x_2)\in T$,
where $T$ is the set of all typical pairs. 

Since we assumed that the processes are physically unrelated, it seems
like the choice of the state of the first process should not change
whatever states are possible for the second one. Therefore, we 
would expect that the set of physically possible (not abnormal) pairs
coincides with the Cartesian product $T=T_1\times T_2$. 

In principle, it is possible to find such a $T$: e.g., we can take a
one-point set $T$. However, as we will show, if we take into
consideration that typical states must form a majority in some
reasonable sense, then such $T=T_1\times T_2$ is no longer possible.
In other words, {\it every two processes in the world are related},
even if the equations that describe these processes are independent.

The strongest result occurs if we consider two {\it identical}
processes, for which it is natural to assume that if $(x_1,x_2)\in T$,
then  $(x_2,x_1)\in T$:
\smallskip

\noindent{\bf Definition 14.} {\sl 
\begin{itemize}
\item Let $X$ be an arbitrary set. By a {\it permutation}, we mean a mapping
$X\times X\to X\times X$ defined as $(x_1,x_2)\to (x_2,x_1)$.
\item For arbitrary sets $X_1\times X_2$, we say that a set
$T\subseteq X_1\times X_2$ is {\it factorizable} if $T=T_1\times
T_2$ for some $T_i\subseteq X_i$.
\item Let $p$ be a probability measure on $X$ that is {\it
non-atomic} (i.e., $p(\{a\})=0$ for all $x\in X$). We say that a set 
$T_i\subseteq X$ is a {\it majority set} if $p(T_i)>1/2$. 
We say that a set $T\subseteq X_1\times X_2$ is a {\it majority set}
if $(p\times p)(T)\ge 1/2$.
\end{itemize}}

\noindent In physical terms, 
factorizable sets correspond to truly independent processes:
\smallskip
\newpage

\noindent {\bf Definition 15.} {\sl 
\begin{itemize}
\item Let $X_1$ and $X_2$ be sets. Elements of
$X_1$ will be called {\it states of the first process}. Elements of
$X_2$ will be called {\it states of the second process}. A pair
$(x_1,x_2)\in X_1\times X_2$ is called a {\it joint state}. 
\item Let $T\subseteq X_1\times X_2$ be a set of joint states. States
from $T$ will be called {\it physically possible.}
\item We say that a state $x_1\in X_1$ of the first process is {\it
physically possible} if $(x_1,x_2)\in T$ for some $x_2\in X_2$.
\item We say that a state $x_2\in X_2$ of the second process is {\it
physically possible} if $(x_1,x_2)\in T$ for some $x_1\in X_1$.
\item Let a state $x_1\in X_1$ be given. We say that {\it given a
state $x_1$, the state $x_2$ is possible for the second process} if
$(x_1,x_2)\in T$. The set of all states of the second process that are
possible for a given $x_1$ will be denoted by $P_2(x_1)$.
\item Similarly, we can define $P_1(x_2)$.
\item We say that the processes are {\it truly independent} if the set
of possible states of the seocnd process does not depend on the state
of the first process (and vice versa), i.e., if $x_1$ and $x_1'$ are
both physically possible, then $P_2(x_1)=P_2(x_1')$. 
\end{itemize}}

\noindent The following result is easy to prove:
\medskip

\noindent
{\bf PROPOSITION 15.} {\it Two processes are truly independent iff
the set $T$ is factorizable.}
\medskip

\noindent
{\bf PROPOSITION 16.} {\it Let $X$ be a definable separable 
metric space, and let $T$ be an infinite 
set of typical elements of $X\times X$
that is invariant w.r.t. permutation. Then, $T$ is not factorizable
(i.e., two processes are not truly independent).}
\medskip

\noindent
The proof of this Proposition is based on the following Lemma that may
be of independent interest:
\medskip

\noindent
{\bf LEMMA 2.} {\it If $X$ and $Y$ are definable sets, $f:X\to Y$ is a
definable mapping, and $T$ is a set of typical elements of $X$, then
$f(X)$ is a set of typical elements of $Y$.}
\medskip

\noindent
In other words, this Lemma says that if an element $x\in X$ is not
abnormal, then its image is also not abnormal.

In the general case, we can usually assume that $X_1=X_2$ (e.g., in
quantum mechanics, the set of all possible states of any system is a
Hilbert space $L^2$).
\medskip

\noindent
{\bf PROPOSITION 17.} {\it Let $X$ be a definable separable 
metric space, and let $T$ be a majority 
set of typical elements of $X_1\times X_2$.
Then, $T$ is not factorizable (i.e., two processes are not truly 
independent).}
\medskip

\noindent
{\bf PROPOSITION 18.} {\it Let $X$ be a definable separable metric
space, and let $T_1$ and $T_2$ be majority sets of typical elements of 
$X$. Then, $T=T_1\times T_2$ is not a set of typical elements of
$X\times X$ (i.e., the processes are not truly independent).}
\medskip

\noindent
The proof of these statements is based on the following Lemma:
\medskip

\noindent
{\bf LEMMA 3.} {\it If $T$ is a set of typical elements of $X$, then
every non-empty subset of $T$ is also a set of typical elements of
$X$.}
\medskip

\noindent
If we have $s>2$ systems, then we can prove an even stronger
statement:
\medskip

\noindent
{\bf PROPOSITION 19.} {\it Let $X$ be a definable separable metric
space, and let $T_1, \ldots , T_s$ be sets of typical elements of 
$X$ for which $p(T_i)>1/s$ for all $i$. Then, $T=T_1\times\ldots \times
T_s$ is not a set of typical elements of
$X\times \ldots \times X$.}
\medskip

\noindent
We can generalize the above definition of true independence to $s>2$
processes, and claim that under the conditions of Proposition 19,
these $s$ processes are not truly independent. 

This proposition take into consideration the fact that 
``typical'' does not necessarily mean ``belonging to 
the majority'': e.g., a ``typical professor'' may combine several
features, each of which may be typical for a majority, but when
combined, may be rather rare.

Let us describe how these results are related to theoretical physics.
\medskip

\noindent {\bf EPR paradox.}
Analyzing quantum mechanics, Einstein, Podolsky and
Rosen came up with the conclusion that in quantum mechanics, it is
potentially possible to have correlation between the states of 
spatially separated particles at the same moment of time. This
conclusion clearly contradicts special relativity, according to which
immediate commununication between spatially spearated events is
impossible. Because of this contadiction, this conclusion was called a
paradox (named EPR by first letters of their
names). For a detailed description and references, see, e.g.,
\cite{Penrose 1989}.
The above results show that if we take into consideration the fact
that a theory is not only equations, but also initial conditions, then
connection even between 
spatially separated events {\it is} possible, so EPR paradox is not a
paradox anymore (other solutions to this paradox are presented in
\cite{Penrose 1989}). 
\medskip

\noindent{\bf Interaction between parallel worlds.} 
Modern physics is formulated in terms of probabilities. Because of
that, even if we measure everuthing accurately, we cannot
uniquely predict the results of future experiments. One way to
describe it is to say that instead of a single world history, there
are several possible world histories. Usually, only one history is
consider real, all others are viewed as purely mathematical objects.
However, strating from Wheeler and Everett, some researchers started
to consider the possibility that all world histories are real: one of
them describes our world, in which we live. Others describe other
worlds (these other worlds do not intersect with ours are are
therefore called {\it parallel worlds} \cite{Penrose 1989}). 

If we assume that parallel
worlds do not influence our world, then whether we call them real or
not is a question of semantics: no experiments in this world are
influenced by anything that happens in these parallel worlds. 
However, some theorists suggested that a small interaction is possible.
For example, in 1972, 
A. Sakharov have suggested that the space-times of the
worlds do have intersections; these intersections are elemenntary
particles, and observable properties of the particles can be
interpreted as topological characteristics of the intersection (for
details and references, see, e.g., \cite{Misner 1973}). 
From the described 
viewpoint, this is a quite natural idea: as soon as
we adopted the model with parallel worlds, then
automatically we adopted the postulate that what's 
going on in all these worlds is typical with respect to
this theory. Hence, due to the above propositions, what is happening
in one of the worlds can influence the others.


\section{Proofs}

\noindent 
{\bf Proof of Proposition 1.} We have already mentioned that there are
denumerably many definable sequences of sets. Therefore, there are no
more than denumerably many sequences of sets $A_n$ for which
$A_n\supseteq A_{n+1}$ and $\cap A_n=\emptyset$. So, we can enumerate
such sequences. Let us denote the elements of the $k-$th
sequence by $A^k_0,A^k_1,\ldots ,A^k_n,..$. For every $k$ and for every
$i$, from $\cap A^k_n=\emptyset$ and monotonicity, it follows that 
$p_i(A^k_n)\to 0$. This means that there exists an integer $N^k_i$ such
that if $n\ge N_i$, then $p_i(A^k_n)\le\varepsilon\cdot 2^{-k}$. Let
us define $N^{(k)}=\max(N^{(k)}_1,\ldots ,N^{(k)}_m)$. Then, $N^{(k)}\ge
N^{(k)}_i$ for all $i$, and therefore, 
$p_i(A^k_{N^{(k)}})\le\varepsilon\cdot 2^{-k}$.
As $T$, we will take the complement to the set 
$$A=\bigcup_{k=1}^\infty A^k_{N^{(k)}}.$$ 
It is easy to see that $A$ is a set of typical elements: indeed, for
every sequence $A^k_0,A^k_1,\ldots ,A^k_n,\ldots $, we have $T\cap
A^k_{N^{(k)}}=\emptyset$. Now, let us show that elements from $A$ are
rare. Indeed, for every $i$, we have $p_i(A)\le \sum p_i(A^k_{N^{(k)}})$. 
For every $i$, we have $p_i(A^k_{N^{(k)}})\le\varepsilon\cdot 2^{-k}$, and 
therefore, $p_i(A)\le\sum (\varepsilon\cdot 2^{-k})=\varepsilon$. Q.E.D.
\smallskip

\noindent
{\bf Proof of Proposition 2.} As $A_n$, let us take the set of all
sequence $r\in S$ for which first $n$ experiments confirm the theory
$\cal T$, but some further experiments do not confirm $\cal T$. Then,
it is easy to show that $A_n\supseteq A_{n+1}$ for all $n$. Since the
theory is physically meaningful, we have 
$\cap A_n=\emptyset$. Therefore, there exists $N$ for which $A_N\cap
T=\emptyset$, i.e., for which all {\it not abnormal} sequences belong
to the complement of $A_N$. Due to our definition of $A_N$, $r\not\in
A_N$ means that if first $N$ experiments confirm the theory, then this
theory is correct. Q.E.D.
\smallskip

\noindent
{\bf Proof of Proposition 3.} As $A_n$, take the set of all pairs
$(\cal T,r)$ for which first $n$ experiments rom the sequence $r$
confirm the theory, but the theory is not correct on $r$. The proof is
similar to the proof of Proposition 2. 
\smallskip

\noindent 
{\bf Proof of Proposition 4.} Let us prove this result for an
arbitrary definable metric space $X$ with a metric $d$. 
As $A_n$, let us take the set of all $x$
for which $0<d(x,a)\le 2^{-n}$. This sequence is decreasing, and $\cap
A_n=\emptyset$. Therefore, there exists an $N$ for which $A_N\cap
T=\emptyset$. This means that none of the elements from $T$ belong to
$A_N$. This, in its turn, means that for elements $x\in T$, either
$d(x,a)=0$, or $d(x,a)>2^{-n}$. For $\varepsilon=2^{-n}$, we get the
desired result. 
\smallskip

\noindent{\bf Proof of Lemma 1.} 
A set $S$ in a metric space $X$ is compact iff it is closed and for every
$\varepsilon$, it has a finite $\varepsilon-$net, i.e., a finite set
$S(\varepsilon)$ with the property that every $s\in S$, there exists
an element $s(\varepsilon)\in S(\varepsilon)$ that is
$\varepsilon-$close to $s$. 

The closure of $T$ is clearly closed, so, 
to prove that the closure of $T$ is a
compact, it is sufficient to prove that it has an $\varepsilon-$set
for all $\varepsilon$. For that, it is
sufficient to prove that for every $\varepsilon>0$, there exists a
$\varepsilon-$net for $T$.
 
If a set $S$ is a $\varepsilon-$net $S(\varepsilon)$, and
$\varepsilon'>\varepsilon$, then, as one can easily see,
 this same set $S(\varepsilon)$ is also a
$\varepsilon'-$net for $S$. Therefore, it is sufficient to show that 
$\varepsilon-$nets for $T$ exist for 
$\varepsilon=2^{-k}, k=0,1,2,\ldots $

Let us fix $\varepsilon=2^{-k}$. 
Since $X$ is definably separable, there exists a 
definable sequence $x_1,\ldots ,x_i,\ldots $ that is everywhere dense in $X$. 
As $A_n$, we will take the complement to the union $U_n$ of $n$ closed
spheres $D_\varepsilon(x_1)$, \ldots , $D_\varepsilon(x_n)$ of radius
$\varepsilon$ with centers in $x_1,\ldots ,x_n$. Clearly, $A_n\supseteq
A_{n+1}$. Since $x_i$ is an everywhere
dense sequence, for every $x$, there exists an $n$ for which $x\in
D_\varepsilon(x_i)$ and for which, therefore, $x\in U_n$ and 
$x\not\in A_n=X\setminus A_n$. Hence,
the intersection of all the sets $A_n$ is empty. Therefore, there
exists an $N$ for which $A_N\cap T=\emptyset$. This means that
$T\subseteq U_N$. This, in its turn, means that the elements
$x_1,\ldots ,x_N$ form an $\varepsilon-$net for $T$.  

So, the set $T$ has an $\varepsilon-$net for $\varepsilon=2^{-k},
k=0,1,2,\ldots $. Hence, $\overline T$ is compact.
Q.E.D.
\smallskip

\noindent
{\bf Proof of Proposition 5.} This proof follows from the known result
that if a function $f$
is continuous and 1-1 on a
compact, then its inverse is also continuous (see, e.g.,
\cite{TA77}). In our case, such a function is $f:\overline T\to
f(\overline T)$. 
\smallskip

\noindent {\bf Proof of Proposition 6.} 
Due to Proposition 5, the inverse function
$f^{-1}$ is continuous on $f(T)$. In particular, 
for every $l$, there exists 
a $\delta>0$ such that if $x,x'\in T$ and 
$d(f(x),f(x'))\le \delta$, then $d(x,x')\le 2^{-l}$. 
If we know $\delta$, then we can compute the desired 
approximation to $x$ as follows. Since $X$ is definably separable, there
exists a definable sequence $x_n$ that is everyweher dense in $X$. Uisng
this sequence, we: 
\begin{itemize}
\item Compute $f(x)$ with accuracy $\delta/8$; the result of this
computation (one of the elements of the everywhere dense sequence $y_m$)
will be denoted by $\tilde f(x)$.
\item For $n=1,2,\ldots $, compute $f(x_n)$ with accuracy $\delta/8$, and
for the result $\tilde f(x_n)$ of this computation, we compute the
distance $d(\tilde f(x),\tilde f(x_n))$ with an accuracy $\delta/8$. 
When this estimate $\tilde d$ is $\le \delta/2$, 
we stop, and produce $x_n$ as the
desired result.  
\end{itemize}
Let us show that this algorithm will work. 
\begin{itemize}
\item First, let us prove that this algorithm will stop. 
Indeed, since the sequence
$x_n$ is everywhere dense in $X$, we have a subsequence $x_{n_k}$ that
tends to $x$. Since $f$ is continuous, we have $f(x_{n_k})\to f(x)$. So,
there exists a $k$ for which $d(f(x_{n_k}), f(x))\le\delta/8$. Since
$\tilde f(x)$ and $\tilde f(x_{n_k})$ are $(\delta/8)-$approximations to
$f(x)$ and $f(x_{n_k})$, we can conclude that 
$d(\tilde f(x_{n_k}),\tilde f(x))\le 
d(\tilde f(x_{n_k}),f(x_{n_k}))+d(f(x_{n_k}),f(x))+d(f(x),\tilde f(x))\le
\delta/8+\delta/8+\delta/8=(3/8)\delta$. Hence $\tilde d\le
d(\tilde f(x_{n_k}),\tilde f(x))+\delta/8\le \delta/2$. 
So, if the algorithm did not
stop before the value $n_k$, it will stop at this point.
\item Let us now show that the algorithm produces the desired value.
Indeed, if $\tilde d\le\delta/2$, then 
$d(\tilde f(x_{n}),\tilde f(x))\le \tilde d+\delta/8\le \delta/2+\delta/8$,
and $d(f(x_{n}),f(x))\le d(\tilde f(x_{n_k}),\tilde f(x))+
d(\tilde f(x_{n}),f(x_{n}))+d(f(x),\tilde f(x))\le
\delta/2+\delta/8+\delta/8+\delta/8<\delta$. Hence, due to our choice of
$\delta$, we have $d(x,x_n)\le 2^{-l}$.
\end{itemize}
So, to complete the description of the algorithm, we must describe how
to compute $\delta$.

We must find $\delta$ such that if $d(f(x),f(x'))\le\delta$, then
$d(x,x')\le 2^{-l}$. To find this $\delta$, let us choose an integer
$p$. Since $f$ is constructively continuous, we can compute the value
$\eta$ such that if $d(x,x')\le\eta$, then $d(f(x),f(x'))\le 2^{-p}$.
Let us take $\beta=\min(\eta,2^{-p})$. For this choice of $\beta$, 
if $d(x,x')\le\beta$, then
$d(x,x')\le 2^{-p}$ and $d(f(x),f(x'))\le 2^{-p}$. Let us find a
$\beta-$net $x^{(1)},\ldots ,x^{(m)}$ 
for $X$. This can be done similarly to the proof of 
Lemma 1, only instead of referring to existence of the desired $N$, we use
the expert to produce such an $N$. 
For this $\beta-$net, we take all pairs $x^{(i)}, x^{(j)}$ for which
$d(x^{(i)},x^{(j)})\ge 2^{-l}-2\beta$, and find the smallest value $M$ of
$d(f(x^{(i)}), f(x^{(j)}))$ for all such pairs. If $M> 2\beta$, then
we return $\delta=M-2\beta$. Else, we increase $p$ by 1, and repeat the
process again and again. 
Let us prove that this part of the algorithm does produce the correct
value of $\delta$ (and thus, that the entire algorithm is correct).
Indeed:
\begin{itemize}
\item Let us first show that this algorithm will stop. Indeed, 
due to Proposition 5, there exists a value $\delta'>0$ for which if
$d(f(x),f(x'))\le\delta'$, then $d(x,x')\le (1/2)\cdot 2^{-l}$. So, if
$d(x^{(i)},x^{(j)})>(1/2)\cdot 2^{-l}$, then
$d(f(x^{(i)}),f(x^{(j)}))>\delta'$. Hence, if we take
$p$ so big that $2^{-p}<\min(\delta'/2,(1/4)\cdot 2^{-l})$, then from
$d(x^{(i)},x^{(j)})\ge 2^{-l}-2\beta$, and from $\beta\le 2^{-p}<
(1/4)\cdot 2^{-l}$, we can conclude that
$d(x^{(i)},x^{(j)})>(1/2)\cdot 2^{-l}$,
and therefore, that $M\ge \delta'>2\cdot 2^{-p}\ge 2\beta$. 
\item Let us now show that if it did stop, then we got the desired
$\delta$. Indeed, let $d(f(x),f(x'))\le M-2\beta$. Since elements 
$x^{(i)}$ are a $\beta-$net for $X$, there exist elements $x^{(i)}$ and
$x^{(j)}$ that are $\beta-$close to $x$ and $x'$ correspondingly. Sue to
the choice of $\beta$, we can conclude that $d(f(x),f(x^{(i)})\le\beta$
and $d(f(x'),f(x^{(j)})\le\beta$. Hence, $d(f(x^{(i)}), f(x^{(j)}))\le
d(f(x),f(x'))+2\beta\le M$. By definition of $M$, this means that
$d(x^{(i)},x^{(j)})\le 2^{-l}-2\beta$. Therefore, $d(x,x')\le
d(x^{(i)},x^{(j)})+2\beta\le 2^{-l}$.
\end{itemize}
So, the second part of the algorithm produces correct $\delta$. Q.E.D.
\smallskip

\noindent
{\bf Proof of Proposition 7.} This proof is similar to the proof of
Proposition 2. Let us fix $M$. As $A_n$, let us take the set of all
testing sequences $s\in S$ for which the method $M$ passes this first $n$
tests, but fails some other test. Then, 
it is easy to show that $A_n\supseteq A_{n+1}$ for all $n$. Since the
testing method is assumed to be complete, we have 
$\cap A_n=\emptyset$. Therefore, there exists $N$ for which $A_N\cap
T=\emptyset$, i.e., for which all {\it not abnormal} testing sequences belong
to the complement of $A_N$. Due to our definition of $A_N$, $s\not\in
A_N$ means that if the method passes first $N$ tests, then this method
is correct. Q.E.D.
\smallskip

\noindent{\bf Proof of Proposition 8.} 
As $A_n$, we will take the 
set of all sequences for which for some $k$, 
$c(x_k,\ldots ,x_{k+d-1})\le 2^{-n}$ 
and $d(x_k,x)>\varepsilon$. Clearly, $A_n\supseteq A_{n+1}$.

Let us show that the intersection $\cap A_n$ is empty. Indeed, suppose
that the sequence $\{x_k\}$ belongs to this intersection. This means
that for every $n$, there exists a $k(n)$ such that 
$c(x_{k(n)},\ldots ,x_{k(n)+d-1})\le 2^{-n}$ and $d(x_{k(n)},x)>\varepsilon$.
If some value $k$ is equal
to $k(n)$ for infinitely many $n$, this means that
$c(x_k,\ldots ,x_{k+d-1})\le 2^{-n}$ for all $n$ and hence, that
$c(x_k,\ldots ,x_{k+d-1})=0$. From the definition of a stopping cruterion,
it then follows that $x_k=x$, so $d(x_k,x)=0\not >\varepsilon$. Hence,
$k(n)\to\infty$, so (since $\{x_k\}$ is convergent), $d(x_{k(n)},x)\to
0$ and $d(x_{k(n)},x)\not >\varepsilon$. The contradiction shows that
the intersection is empty.

So, there exists an $N$ for which $A_N\cap T=\emptyset$. Hence, we can
take $\delta=2^{-N}$. Q.E.D.
\smallskip

\noindent
{\bf Proof of Proposition 9.} As $A_n$, we take the set of all
computations $u$ for which $t(u)>n$ and $t(u)\ne\infty$. Then, there
exists $N$ such that $A_N\cap T=\emptyset$. Hence, we can take $N$ as
the desired $T_0$. 
\smallskip

\noindent
{\bf Proof of Proposition 10.} As $A_n$, we take the set of all
polynomial-time algorithms $\cal U$ for which $t_{\cal U}(x)>n\cdot
({\rm len}(x))^n$ for some input $x$. Clearly, $A_n\supseteq A_{n+1}$. 
Let us prove that $\cap A_n=\emptyset$. 

Indeed, let us take an arbitrary algorithm $\cal U$ from $U$ and show
that it is does not belong to the intersection $\cap A_n$. Every algorithm
from $\cal U$ is time-polynomial, i.e., 
$t_{\cal U}(x)\le c\cdot ({\rm len}(x))^k$ for some $c$ and $k$.
Therefore, for $n=\max(c,k)$, we have ${\cal U}\not\in A_n$. Hence,
${\cal U}\not\in \cap A_n$, and therefore, $\cap A_n=\emptyset$. 

So, there exists an $N$ for which $A_N\cap T=\emptyset$. This means
that if ${\cal U}\in T$, then 
$t_{\cal U}(x)\le n\cdot ({\rm len}(x))^n$ for all inputs $x$. So, we
can take $C=K=N$. Q.E.D.
\smallskip

\noindent
{\bf Proof of Proposition 11.} As $A_n$, we take $\{u\,|\,|f(u)|>n\}$.
Then, $A_n\supseteq A_{n+1}$, $\cap A_n=\emptyset$, and hence, there
exists an $N$ for which $A_N\cap T=\emptyset$. This means that if
$u\in T$, then $|f(u)|\le N$. Q.E.D.
\smallskip

\noindent
{\bf Proof of Proposition 12.} Take $A_n=\{u\,|\,f(u)\ne 0\&
|f(u)|\le 2^{-n}\}$. Then, $A_n\supseteq A_{n+1}$, $\cap
A_n=\emptyset$, and so, there exists an $N$ for which 
$A_N\cap T=\emptyset$. So, we can take $\varepsilon=2^{-N}$. Q.E.D.
\smallskip

\noindent
{\bf Proof of Proposition 13.} Let $s_1, s_2\in T$ and $s_1\ne s_2$.
Then, $d(s_1,s_2)>0$. Since $X$ is definably separable, there
exist a definable everywhere dense sequence $x_n$. In particular, there
exists an $n$ for which $d(s_1,x_n)<(1/2)\cdot d(s_1,s_2)$. From the trangle
inequality, it easily follows that $d(s_2,x_n)\ge d(s_1,s_2)-d(s_1,x_n)
>(1/2)\cdot d(s_1,s_2)$. So, $d(s_2,x_n)>d(s_1,x_n)$. On the interval
$[d(s,x_1),d(s_2,x_n)]$, there exists a rational (and hence definable)
point $r$. 

Let us take $A_n=\{x\,|\,|d(x,x_n)-r|\le
2^{-n}\,\&\,d(x,x_n)\ne 0\}$. This is a decreasing sequence, and $\cap
A_n=\emptyset$, so, there exists an $N$ for which $A_N\cap T=\emptyset$.
This means that if $x\in T$, then either $d(x,x_n)<r-2^{-N}$, or
$d(x,x_n)=r$, or $d(x,x_n)>r+2^{-N}$. 
So, we can take $S_1=\{x\,|\,d(x,x_n)<r-(1/2)\cdot 2^{-N}\}$, and  
$S_2=\{x\,|\,d(x,x_n)>r-(1/4)\cdot 2^{-N}\}$.
Both sets are open, $S_1\cap S_2=\emptyset$, $s_i\in S_i$, and
$T\subseteq S_1\cup S_2$. So, $T$ is really completely disconnected.
Q.E.D. 
\smallskip

\noindent
{\bf Proof of Proposition 14.} Since $G$ is non-trivial, there exist an
element $x\in T$ and $g\in G$ for which $a_g(x)\ne x$. 
For this $x\in T$, the orbit
$Gx=\{a_g(x)\,|\,g\in G\}$ is a continuous image of the connected set
$G$ and is, therefore, connected. It contains more than 2 points.
Since $T$ is completley disconneted, $T$ cannot contain a connected
subset different from a single point. Hence, $Gx\not\subseteq T$. This
means that there exists a $g$ for which $a_g(x)\not\in T$. So,
$x\in T$, $a_g(x)\not\in T$, hence, $a_g(T)\ne T$. Q.E.D.
\smallskip

\noindent
{\bf Proof of Lemma 2.} Indeed, let $A_n$ be a sequence of subsets of
$Y$ for which $A_n\supseteq A_{n+1}$, and $\cap A_n=\emptyset$. Then,
for $B_n=f^{-1}(A_n)$, we have $B_n\supseteq B_{n+1}$. 

If $x\in \cap
B_n$, then $f(x)\in A_n$ for all $n$, so, $f(x)\in\cap A_n$, which
contradicts to our assumption that $\cap A_n=\emptyset$. Hence, $\cap
B_n=\emptyset$. Since $T$ is a set of typical elements, we can
conclude that there exists an $N$ for which 
$B_N\cap T=f^{-1}(A_N)\cap T=\emptyset$. 

Let us show that $A_N\cap f(T)=\emptyset$. Indeed, suppose that there
exists an element $y\in A_N\cap f(T)$. Since $y\in f(T)$, there exists
an $x\in T$ for which $y=f(x)$. This $x$ belongs both to $T$ and to
$f^{-1}(A_N)=B_N$, which contradicts to our choice of $N$. So, such an
element $y$ is impossible. Hence, $A_N\cap f(T)=\emptyset$. Q.E.D. 
\smallskip

\noindent
{\bf Proof of Proposition 16.} We will prove this Proposition by
reduction to a contradiction. Assume that $T=T_1\times T_2$ is a
set of typical elements of $X\times X$ that is invariant w.r.t.
permutations. Since $T$ is invariant w.r.t. permutations, we have
$T_1=T_2$. 

If we take $A_n=\{(x_1,x_2)\,|\,d(x_1,x_2)<2^{-n}\,\&|, x_1\ne x_2\}$,
then we can conclude that there exists an $N$ for which $A_N\cap
T=\emptyset$. This means that if $(x_1,x_2)\in T$ and $x_1\ne x_2$,
then $d(x_1,x_2)\ge 2^{-n}$. Since $T=T_1\times T_2$ and $T_1=T_2$, we can 
reformulate this condition as follows: if $x_1,x_2\in T_1$, and
$x_1\ne x_2$, then $d(x_1,x_2)\ge 2^{-N}$. So, every two elements from
$T_1$ are $\ge 2^{-N}-$different from each other. 

The set $T_i$ is a projection of $T$ on $X$: $T_1=\pi_1(T)$, where $pi_1:
X\times X\to X$ is a definable mapping (defined as $(x_1,x_2)\to
x_1$). So, due to Lemma 2, $T_1$ is a set of typical elements of $X$.

Due to Lemma 1, this set $T_1$ is pre-compact. In a compact set, there
can be at most finitely many elements that are $2^{-N}-$different
from each other. So, $T_1$ is finite. Hence, $T=T_1\times T_1$ is also
finite, and this contradicts to our assumption that $T$ is infinite.
Q.E.D.
\smallskip

\noindent 
{\bf Proof of Lemma 3:} this statement immediately follows from 
Definition 3.
\smallskip

\noindent
{\bf Proof of Proposition 18.} 
We will prove this proposition by reduction to a contradiction. 
Assume that $T_1\times T_2$ is a set of typical elements of $X\times
X$. Similarly to Proposition 16, we can prove that there exists a 
$\varepsilon>0$ such that
for all typical pairs $(x_1,x_2)\in T$, either $x_1=x_2$, or
$d(x_1,x_2)\ge \varepsilon$. 

Due to Lemma 3, the intersection $T_1\cap T_2\subseteq T_1$ 
is a set of all typical elements of $X$. Similarly to the proof of
Proposition 16, we can now conclude that this intersection is finite.
Since $p$ is a non-atomic measure, we have $p(T_1\cap T_2)=0$.
Hence, $p(T_1\cup T_2)=p(T_1)+p(T_2)-p(T_1\cap T_2)=p(T_1)+p(T_2)$.
But $p(T_i)>1/2$, so, $p(T_1\cup T_2)>1$, which contradicts to the
fact that $p$ is a probability measure (and so, $p(T_1\cup T_2)\le
p(X)=1$). The contradiction proves that our assumption is wrong, and
$T$ is not the set of typical elements. Q.E.D.
\smallskip

\noindent
{\bf Proof of Proposition 17.} Assume that $T$ is factorizable, i.e.,
$T=T_1\times T_2$. Then,
due to Lemma 2, each of the sets $T_i$ is a set of typical elements.
Since $T_1\times T_2$ is $(p\times p)-$measurable, 
both its projections $T_i$ 
must be $p-$measurable sets, and $(p\times p)(T_1\times
T_2)=p(T_1)\cdot p(T_2)$. Since $p(T_i)\le 1$, we have $p(T_i)\ge
(p\times p)(T_1\times T_2)>1/2$. So, both sets $T_i$ are majority
sets. The result follows from Proposition 18. Q.E.D.
\smallskip

\noindent 
{\bf Proof of Proposition 19.} This proof is similar to the proof of
Proposition 17. Indeed, suppose that $T_1\times \ldots \times T_s$ is a set
of typical elements for $X\times \ldots \times X$. Then, similarly to that
proof, we conclude that the intersection of each
pair $T_i$ and $T_j$ is finite and therefore, $p(T_i\cap
T_j)=\emptyset$. Hence, $p(T_1\cup \ldots \cup
T_s)=p(T_1)+\ldots +p(T_s)>(1/s)+\ldots +(1/s)=1$, which contradicts to the
assumption that $p$ is a probability measure. Q.E.D.

\subsection*{Acknowledgments}
This work was supported in part by NASA under cooperative 
agreement NCC5-209 and grant NCC 2-1232, by the 
Future Aerospace Science and Technology Program (FAST) 
Center for Structural Integrity of Aerospace Systems,
effort sponsored by the Air Force Office of Scientific Research, Air Force
Materiel Command, USAF, under grant number F49620-00-1-0365, by the 
Grant No. W-00016 
from the U.S.-Czech Science and Technology Joint Fund, and by 
Grant NSF 9710940 Mexico/Conacyt.

\end{document}